# UV/Ozone treatment to reduce metal-graphene contact resistance

Wei Li[1,2], Yiran Liang[1], Dangmin Yu[1], Lianmao Peng[1], Kurt P. Pernstich[2], Tian Shen[2], A. R. Hight Walker[2], Guangjun Cheng[2], Christina A. Hacker[2], Curt A. Richter[2], Qiliang Li[3], David J. Gundlach[2,*], and Xuelei Liang[1,*]

1. Key Laboratory for the Physics and Chemistry of Nanodevices and Department of Electronics, Peking University, Beijing, 100871, China
2. Physical Measurements Laboratory, National Institute of Standards and Technology, Gaithersburg, MD 20899, USA
3. Department of Electrical and Computer Engineering, George Mason University, Fairfax, VA 22030, USA

E-mail:  liangxl@pku.edu.cn; david.gundlach@nist.gov

We report reduced contact resistance of single-layer graphene devices by using ultraviolet ozone (UVO) treatment to modify the metal/graphene contact interface. The devices were fabricated from mechanically transferred, chemical vapor deposition (CVD) grown, single layer graphene. UVO treatment of graphene in the contact regions as defined by photolithography and prior to metal deposition was found to reduce interface contamination originating from incomplete removal of poly(methyl methacrylate) (PMMA) and photoresist. Our control experiment shows that exposure times up to 10 minutes did not introduce significant disorder in the graphene as characterized by Raman spectroscopy. By using the described approach, contact resistance of less than 200 Ω μm was achieved, while not significantly altering the electrical properties of the graphene channel region of devices.

Graphene is considered a candidate material for post-silicon electronics[1], and graphene-based electronic and optoelectronic devices have developed rapidly in recent years[2]. In graphene-based devices, the metal/graphene contact is viewed as a limiting factor in its performance[3]. Ohmic metal/graphene contacts with low contact resistance ($R_c$) are necessary for graphene FET performance to approach its expected high intrinsic speed. To date, the best reported $R_c$ for lithographically defined contacts deposited onto exfoliated graphene flakes ranges from 200 Ω μm to 500 Ω μm.[3, 4] $R_C$ for contacts formed to epitaxial graphene on SiC have been reported to be less than 100 Ω μm[5] and with specific contact resistivity ($\rho_c$) of order $10^{-7}$ Ω $cm^2$.[5, 6] Reported values of $R_c$ for chemical vapor deposited (CVD) graphene typically range from 500 Ω μm to 1000 Ω μm.[7, 8] Despite the technological attractiveness of CVD grown graphene, these contact resistances remain too large for most applications and are far from that reported for contacts to epitaxial graphene on SiC.

Possible contributors to the large, experimentally-determined metal/graphene contact resistance include: dipole formation at the interface due to charge transfer, perturbation of the graphene beneath the metal[9], and contamination of the metal/graphene interface[10]. Interface contamination during the device fabrication, especially when a photolithography process is employed, is known to be problematic and efforts are made to clean the contact interface in conventional semiconductor processing. Using the same or similar photolithographic processes to pattern contacts onto graphene make it reasonable to expect photoresist residue contamination on graphene surface. Several researchers have introduced methods to reduce the contact resistance. For example, specific contact resistivity as low as $10^{-7}$ Ω $cm^2$ was obtained from short channel length (L < 3 um) transfer length test structures (TLM) on epitaxial graphene by a low power plasma treatment.[6] Unfortunately, the plasma treatment is aggressive and after tens of seconds of treatment the graphene can be seriously degraded, leading to a high variance in device to device contact resistance.[11] $R_c$ less than 100 Ω μm and specific contact resistivity less than $10^{-7}$ Ω $cm^2$ have been reported for contacts formed to epitaxial graphene on SiC by other researchers, but details about device processing, importantly, contact formation are absent in the report.[5] Using a double contact device geometry $R_c$ of 200 Ω μm to 500 Ω μm on CVD graphene was reported.[12] Similar contact resistance was reported for metal contacts to CVD graphene by introducing an Al sacrificial layer.[13] Such strategies complicate the device fabrication process and make it necessary to develop a simple and robust process for reducing the metal/graphene contact resistance.

In this work we report significantly reduced contact resistance to CVD, single-layer graphene obtained by using an easily graphene surface cleaning method: ultraviolet ozone (UVO) treatment. UVO is a common cleaning process used in semiconductor device research and manufacturing, and in applications requiring critically clean interfaces such as those involving the assembly of molecules on metal or oxide surfaces for which aggressive plasma and ion bombardment processes cannot be tolerated.[14] By using UVO, we are able to reduce $R_c$ to mechanically-transferred, CVD single-layer graphene to less than 200 Ω μm while preserving the electrical properties of the graphene device.

In this study we fabricated transfer length method (TLM) test structures from single layer graphene that was grown on Cu foil by CVD and then mechanically transferred onto a heavily doped Si substrate with 300 nm $SiO_2$ using a "modified RCA clean method".[15] Following the graphene transfer onto the $SiO_2$ surface, the test structures were fabricated by using conventional contact photolithography and metal deposition. The process flow is shown schematically in Fig. 1. After opening the windows in the photoresist layer the substrate was placed into a commercial UVO system to remove resist residue prior to metallization. Ti (20 nm) / Au (80 nm) was evaporated and patterned by lift-off process. A second photolithography step and oxygen plasma etching were used to pattern the graphene channel. The sacrificial photoresist mask used to protect the graphene channel region during the etch process was removed by using solvents.

We first evaluate the aggressiveness and effectiveness of the UVO cleaning step by using atomic force microscopy (AFM) (Fig. 2) and Raman spectroscopy (Fig. 3). A PMMA layer was used as a polymer support layer during the mechanical transfer process of the graphene and needed to be removed at the end of the transfer process before proceeding with the first photolithographic step. However, PMMA was not thoroughly removed with solvents for overnight immersion and a thin residue layer still remained on the graphene surface. This is visible in Fig. 2a, which shows the AFM topography image for a transferred, single-layer graphene domain after the solvent removal of PMMA. Next, a commercial photoresist was spin coated onto the transferred graphene surface and the substrate went through the same exposure and developing steps used in the fabrication of the TLM test structures. Thus, Fig. 2b shows representative surface topography of the contact regions after developing the resist openings and just prior to the metal deposition. The rough surface features on the single layer graphene indicates substantial resist residue remains on the graphene surface and we expect, in the absence of additional cleaning processes, that this residual resist will prevent the formation of an intimate metal/graphene contact interface. Fig. 2c-f sequentially shows the results of accumulative 5, 10, 16, 22 minutes UVO treatments. The surface appears smooth after about 16 minutes and no further change of the surface topography was observed for UVO exposure up to 22 minutes.

Fig. 3 shows the Raman spectra taken at nominally the same position on the graphene domain as shown in Fig. 2 and in parallel with the AFM topography scans during the series of accumulative UVO treatment times. Raman spectroscopy has become a widely employed characterization method for evaluating the quality of graphene.[16,17] Usually three Raman peaks near 1580 cm$^{-1}$ (G peak), 2650 cm$^{-1}$ (G′ peak), and 1350 cm$^{-1}$ (D peak), are observed in the spectra of graphene.[16] A high D-to-G peak intensity ratio correlates to a greater degree of disorder in the graphene structure and increased charge carrier scattering. In our experiments, the D peak intensity remains relatively low, and a significant change is not observed until 22 minutes of UVO treatment.

It is important to acknowledge here that UVO processes can vary greatly among UVO systems, and depend on the specific configuration and use in individual laboratories (e.g. exhaust rate, feed gas, exposure time, sample-to-grid lamp distance, sample temperature, and lamp intensity). In fact, one early study using aggressive UVO processing conditions reports significant damage to pristine graphene at short time scales.[18] We have collected additional

Raman spectra (not shown) on mechanically-transferred, CVD single-layer graphene post photoresist processing and after UVO treatment in a different UVO system and obtained D-to-G peak intensity ratios similar to that shown in Fig. 3 for the first 10 minutes of UVO treatment. However, a pronounced increase in D peak intensity (substantial increase in D-to-G peak intensity ratio) is observed after 16 minutes of UVO treatment. Additionally, results from preliminary X-ray photoelectron spectroscopy (XPS) studies (not shown) on these same samples reveal changes in the C 1s and Si 2p peak intensities with UVO exposure time that indicate organic contamination and removal with UVO exposure. XPS data indicated the eventual degradation of the graphene when exposed to longer UVO treatment times entirely consistent and coincident with the pronounced emergence of the D peak in the Raman spectral.

Contact resistance was extracted from TLM test structures that were fabricated by using the process flow depicted in Fig. 1. Fig. 4a shows the optical micrograph (contrast enhanced) of a TLM test structure. The width of the photolithographically defined graphene strip (device channel) is 10 μm and width of the metal contacting the graphene strip is 6 μm. Fig. 4b shows a plot of representative width normalized current-voltage (I-V) characteristics of three devices with the same inter-electrode separation (L = 22.5 μm), but with different UVO treatment time, as well as the I-V characteristics for a device with identical L but processed without UVO treatment. All I-V characteristics are linear over a large applied voltage range and indicate the contacts are ohmic. The measurements were taken at room temperature in air. No back-gate voltage was applied to the heavily-doped silicon substrate. For our Ti/Au contacted test structures the neutrality point is shifted positive by many 10's of volts in air and $R_c$ measured far from the Dirac point has been shown to be almost independent of the gate bias.[4]

Fig. 5a shows the measured resistance (combined pad, contact, and graphene channel) versus contact separation of typical TLM structures as a function of contact interface conditioning. The measured resistance is the aggregate value calculated from the linear I-V characteristics for large voltage sweeps (0 V to 0.5 V, 0.01 V steps). Two notable observations are: 1) the total resistance is greatly reduced by the UVO treatment and 2) the change in resistance with L (contact separation or channel length) for test structures with and without UVO treatment is similar. These observations provide a first indication that the contact resistance is strongly affected by the UVO treatment but the channel resistance is not. The contact resistance and the channel resistance were extracted from a linear fit to the data for L > 5 μm and the width normalized contact resistance and graphene sheet resistance are plotted in Fig. 5b. The contact resistance was reduced more than 2 orders magnitude for a 25 minute UVO treatment. As alluded to by the AFM and Raman studies discussed above, even a 10 minute UVO treatment was found to remove enough residue to improve contact formation between the graphene surface and metal, as substantiated by the nearly 100x reduction in the width normalized contact resistance. By using a UVO treatment, we obtained $R_c$ as low as 184 Ω μm (not corrected for the pad resistance), which is to the best of our knowledge the lowest reported normalized contact resistance to CVD single layer graphene. This corresponds to a specific contact resistivity of order $10^{-7}$ Ω cm$^2$ if one assumes a conservative value of 200 nm for the contact transfer length.[19,20] We note that our linear extrapolations of $R_C$ were limited to data collected for devices with L > 5 μm, but we have included data points for devices with L = 5 μm for completeness. Data for short L was excluded from the extrapolation because we consistently observed pronounced deviations from a linear fit to the data at shorter L. This observation is consistent with the reported observed changes in the electrical properties of graphene devices for L < 5 μm that have been ascribed to L approaching the "quasi-ballistic limit" for graphene,[21] and with reports of "negative contact resistance" extracted by TLM for short L devices.[22]

Importantly, during the UVO treatment of the contact region the graphene channel of the device remained masked by the photoresist. From the nearly unchanged values for sheet resistance we conclude that the channel properties of the devices are not greatly affected by our contact treatment method. For completeness, we have characterized the room temperature field-effect properties of the TLM test structures by applying a voltage to the heavily doped substrate w.r.t. the source contact while measuring the drain current. The field effect characteristics were measured under high vacuum (measurement conditions where the neutral point is restored to within a few volts from $V_{GS}=0$). The mobility was extract by a fit to the data using Eq. 1 of Ref. [23] for $V_{DS} = 0.1$ V, and $V_{GS}$ swept from -60 V to 60 V. The mobility averaged over all channel lengths was 1773±574 cm$^2$/Vs, 3264± 32 cm$^2$/Vs, 2178±178 cm$^2$/Vs, and 1725±383 cm$^2$/Vs, for no UVO treatment, 10, 16, and 25 minute UVO treatments, respectively. The average mobility is found to be largely independent of the UVO treatment and we ascribe any variations in the mobility to the polycrystalline nature of the CVD single layer graphene itself and structural imperfection introduced during the mechanical transfer of the graphene.

We have determined through AFM, Raman, and preliminary XPS studies that a major contributor to high contact resistance and poor device reproducibility of CVD, single-layer graphene devices is the resist residue left on the graphene surface after photolithographic processing. UVO is demonstrated to be a convenient and useful process for removing interfacial contamination from graphene and reducing contact resistance to record low values (< 200 Ω μm) for photolithographically-defined, metal contacts deposited onto CVD monolayer graphene. Moreover, the channel properties of graphene devices are not significantly degraded at the expense of improved contact properties. These results contribute to increasing the likelihood that technologically relevant, CVD grown, single-layer graphene will find use in commercial electronic device applications.

**Acknowledgement:**

This work is supported by the Ministry of Science and Technology of China (Grant No. 2011CB921904) and National Natural Science Foundation of China (Grant No. 60971003). Wei Li is partly supported by National Institute of Standard and Technology.

# Figure captions

Figure 1.  Schematic illustration of the graphene device fabrication process after mechanical transfer of the CVD grown single layer graphene and the solvent removal of the sacrificial PMMA layer.

Figure 2.  AFM images of the graphene surface topography throughout the UVO treatment process. (a) after transfer and solvent removal of PMMA, (b) after photolithography, (c)-(f) UVO treatment for 5, 10, 16, 22 minutes, respectively.
Scale bar:1 μm, Color scale: 10 nm.

Figure 3.  Raman spectra series for the transferred CVD single layer graphene during the UVO treatment process. All spectra were taken from roughly the same position for comparison.

Figure 4.  (a) Optical micrograph of a completed graphene TLM test structure. (b) Width normalized electrical characteristics of 22.5μm channel length devices with and without UVO treatment.

Figure 5.  Electrical characteristics of graphene devices. (a) Total resistance vs. contact separation for different TLM test structures without and with UVO treatment. (b) Width normalized contact resistance and graphene sheet resistance with and without UVO treatment.

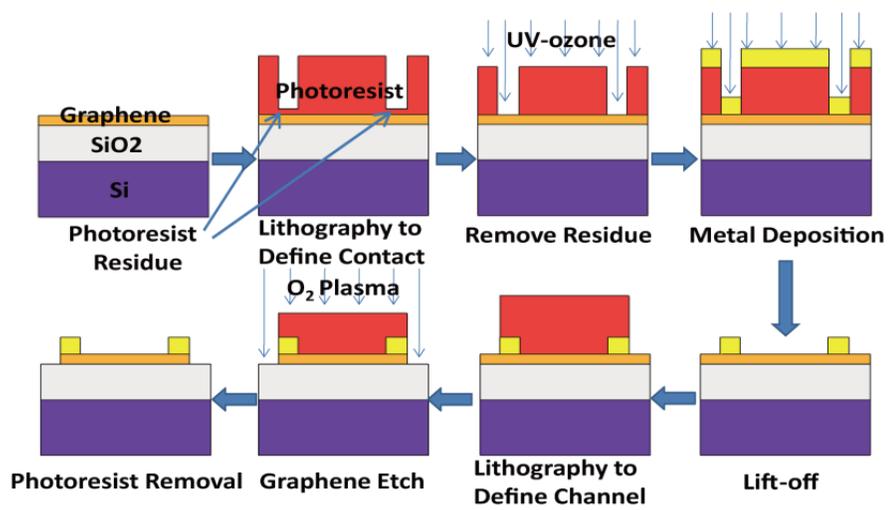

Figure 1 of 5, W. Li, et al.

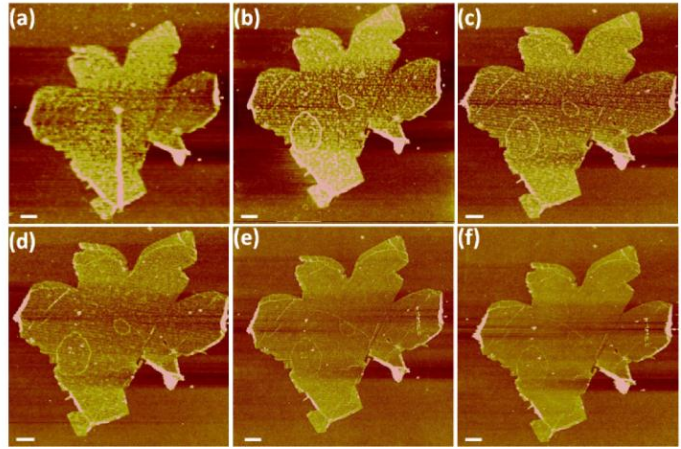

Figure 2 of 5, W. Li, et al.

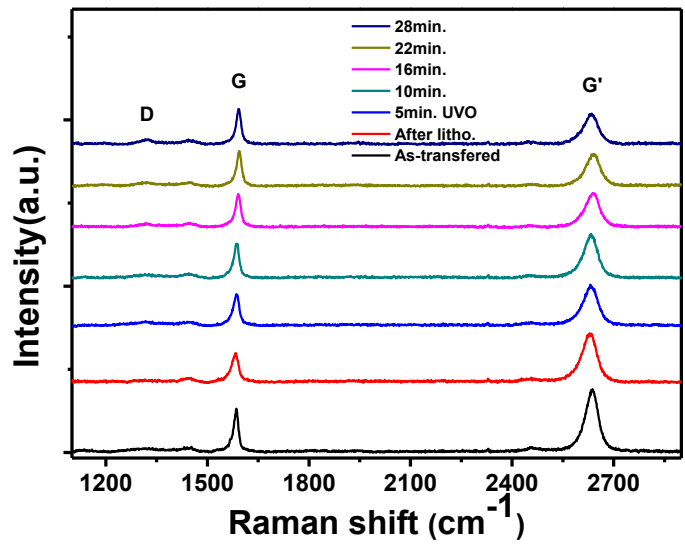

Figure 3 of 5, W. Li, et al.

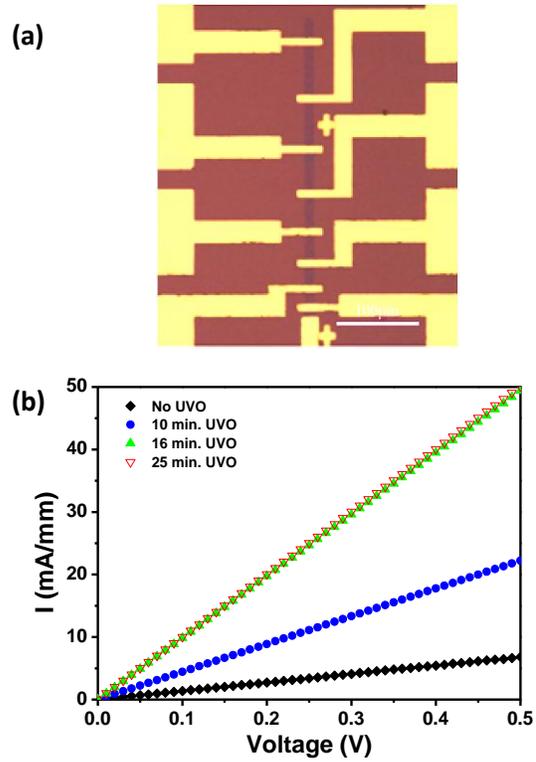

Figure 4 of 5, W. Li, et al.

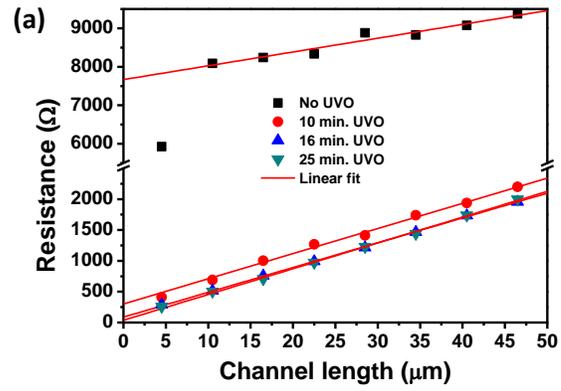

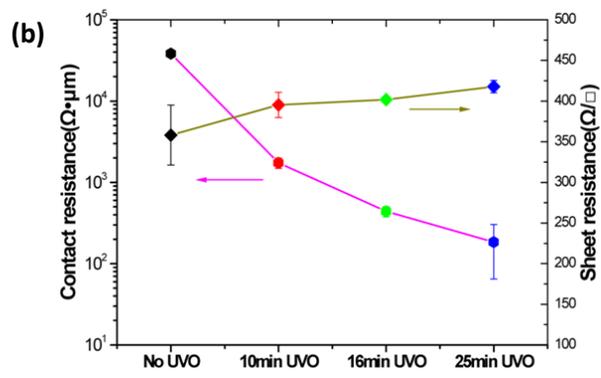

Figure 5 of 5, W. Li, et al.